%% file: LOS_arXiv.tex
\documentclass[12pt]{paper}
\usepackage{latexsym,amssymb,amsfonts,amsmath}
\usepackage{moreverb}
\usepackage{graphicx}
\usepackage{setspace}
\usepackage{multirow}
\usepackage{arydshln}
\usepackage{pgfplots}
\usepackage{booktabs}
\usepackage{pgfplotstable}
\usepackage{bm}
\usepackage[nolists, tablesfirst]{endfloat}
\usepackage{dcolumn}
\usepackage{subcaption}
\usepackage{lscape}
\usepackage[sort&compress,numbers]{natbib}
\makeatletter 
\renewcommand\@biblabel[1]{#1.} 
\makeatother %
\usepackage[colorlinks,bookmarksopen,bookmarksnumbered,citecolor=blue,urlcolor=red]{hyperref}

\input{mytikz}

\begin{document}


\title{Modelling Hospital length of stay using convolutive mixtures distributions}

\author{Adrien Ickowicz, Ross Sparks}





\maketitle

\begin{abstract}
Length of hospital stay (LOS) is an important indicator of the hospital activity and management of health care. The skewness in the distribution of LOS poses problems in statistical modelling because it fails to adequately follow the usual traditional distribution such as the log-normal distribution. The aim of this work is to model the variable LOS using the convolution of two distributions; a technique well known in the signal processing community. The specificity of that model is that the variable of interest is considered to be the resulting sum of two random variables with different distributions. One of the variables will feature the patient-related factors in terms their need to recover from their admission condition, while the other models the hospital management process such as the discharging process. Two estimation procedures are proposed. One is the classical maximum likelihood, while the other relates to the expectation maximisation algorithm. We will present some results obtained by applying this model to a set of real data from a group of hospitals in Victoria (Australia).
\end{abstract}


\maketitle

\section{Motivation of the study}

Length of stay (LOS) is an easily available indicator of hospital activity. It is used for various purposes, such as management of hospital care, quality control, appropriateness of hospital use and hospital planning. LOS is an indirect estimator of resources consumption and of the efficiency of one of the aspects of hospital patient care: bed management. As such a quantity of interest, many works have flourished with the aim of accurately estimate the LOS. Amongst these works, most start with the diagnosis-related group (DRG), that determines the amount of payment allocated to the hospital. The DRGs provide a classification system of episodes of hospitalization with clinically recognized definitions, where it is expected that patients in the same class consume similar quantities of resources, as a result of a process of similar hospital care. The mean of the LOS is used as an indicator of the consumption of resources because of its availability and good relation with raised costs. Hence, we may say that DRGs have been partially created in order to get homogeneous groups with respect to the consumption of services and costs, closely related to LOS.\\
The empirical distribution of LOS is well established to be positively skewed, multi-modal, to contain outliers and to significantly vary between DRGs \cite{Shachtman1986, Marazzi1998}. This heterogeneity of LOS poses a problem for statistical analysis, limiting the use of inference techniques based on normality assumptions. Since a large number of DRGs must be analyzed routinely, automatic procedures are needed for conveniently treating skewness. Different transformations (e.g. the logarithmic one) of LOS have been attempted to attain normality, and subsequently to apply the corresponding tests \cite{Shachtman1986}, etc.). However, these approaches rely on the unrealistic homogeneity assumption on the entire sample. Marazzi et al. \cite{Marazzi1998} assessed the adequacy of three conventional parametric models, Lognormal (long-tailed), Weibull and Gamma (short-tailed), for describing the LOS distribution. But, as Lee et al. \cite{Lee2001} point out, none of them seemed to fit satisfactorily in a wide variety of samples. The main issue is that the assumption of heterogeneous sub-populations would be more appropriate than single DRG populations.\\
 Mixture distribution analysis can clarify whether or not a skewed distribution is composed of heterogeneous components \cite{Quantin1999}. Recently, Atienza \cite{Atienza2008} tried to model the LOS within DRGs using a mixture of Gamma, Weibull and Lognormal distributions. The proposed model shows good fitting properties but requires a complex EM algorithm that will be extremely sensitive to initial values. Carter and Potts \cite{Carter2014} use a purely negative binomial regression to fit their data on knee replacement. They display a relative efficiency of $75\%$ for a very specific LoS of $4-6$ days, the margin error being $2$ days. The results for other LoS were however more disappointing.\\
All these models suffers from various limitations. The idea of fitting a mixture model has several drawbacks. It is computationally complex, especially when the number of mixtures is unknown. It lacks of natural interpretation, in particular is the modes of the mixture do not correspond to a simple combination of explanatory variables. Moreover, all these works failed to describe the exact behaviour of the LOS. 
\begin{figure}[t]
\includegraphics[scale=0.35]{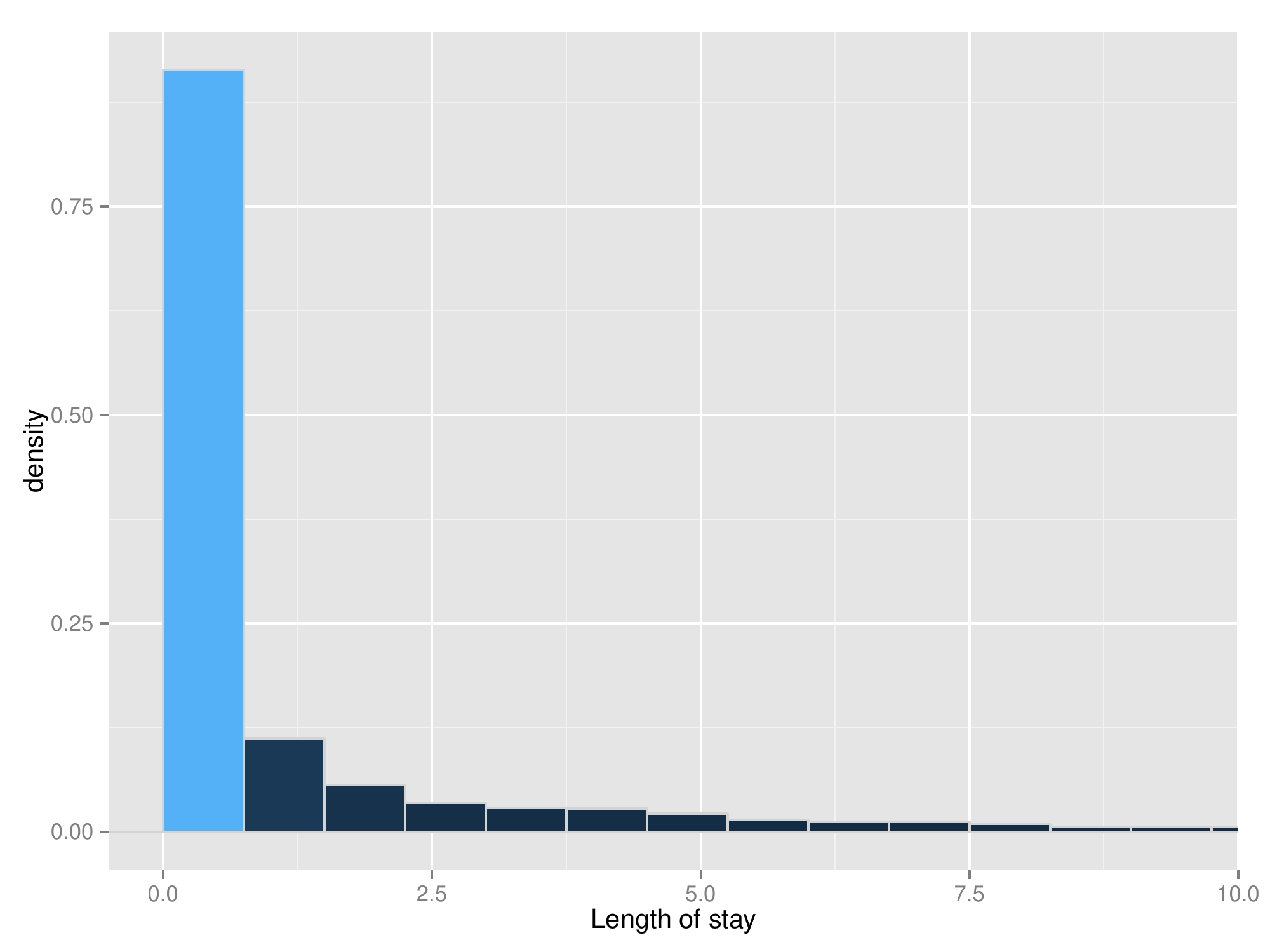}
\includegraphics[scale=0.35]{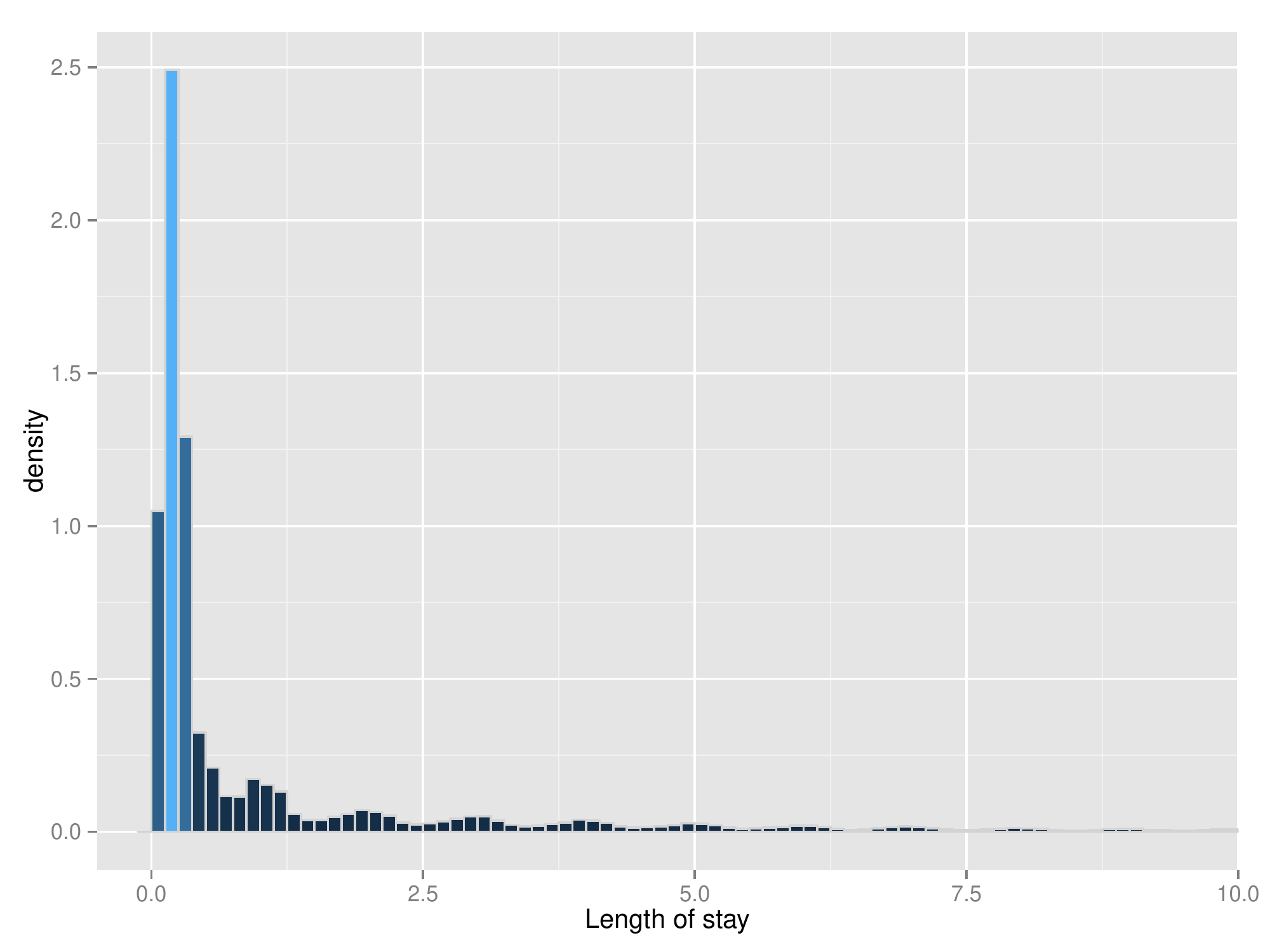}
\caption{\label{fig:hist} Comparison between two histograms of the same data. On the left, break points are every day. On the right, break points are every $6$ hours.}
\end{figure}
Indeed, Figure \ref{fig:hist} presents two histograms of the same LOS data, but with different breakpoints. On the left, the breakpoints are taken every days, while they were taken every $6$ hours on the right histogram.\\
\begin{figure}[t]
\includegraphics[scale=0.23]{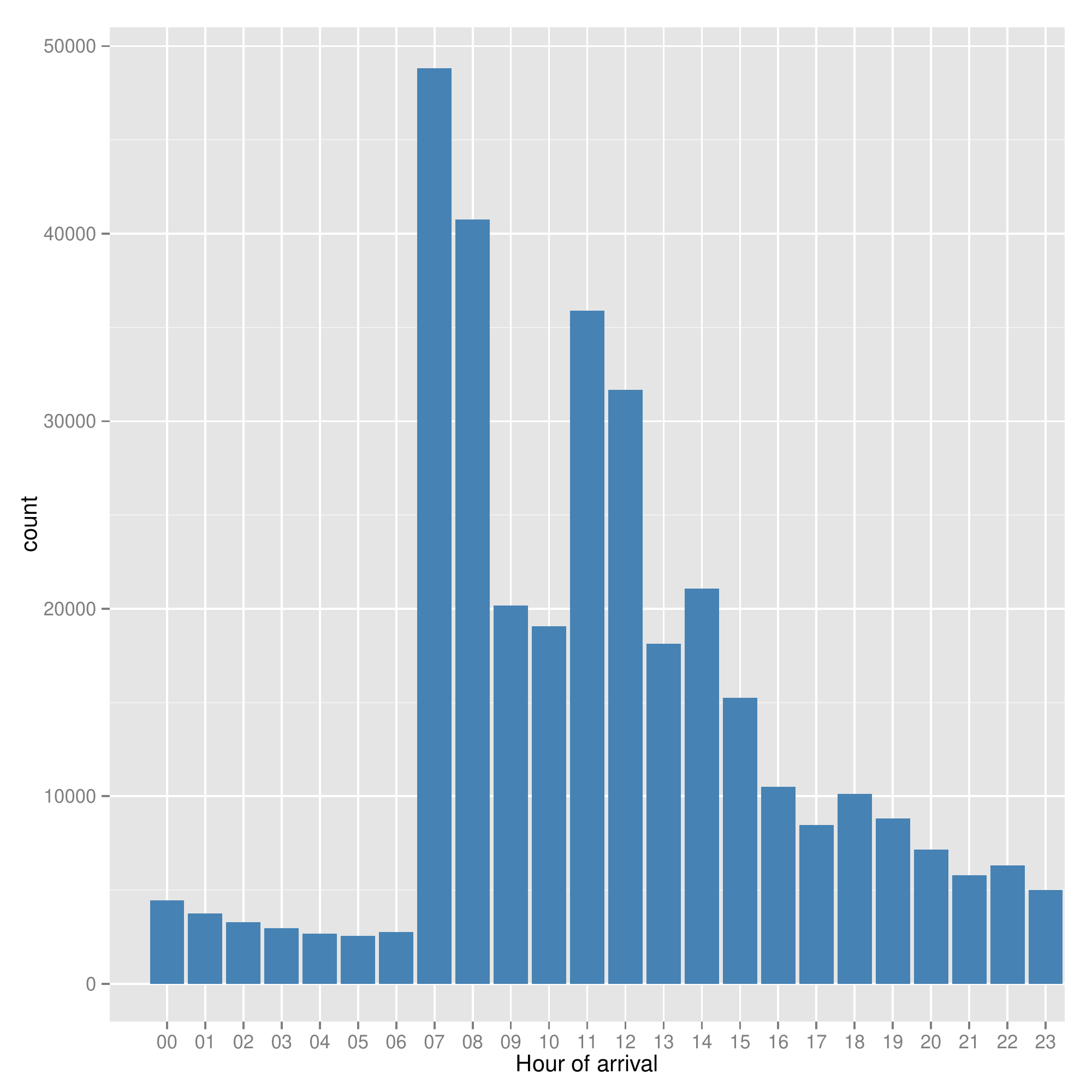}
\includegraphics[scale=0.23]{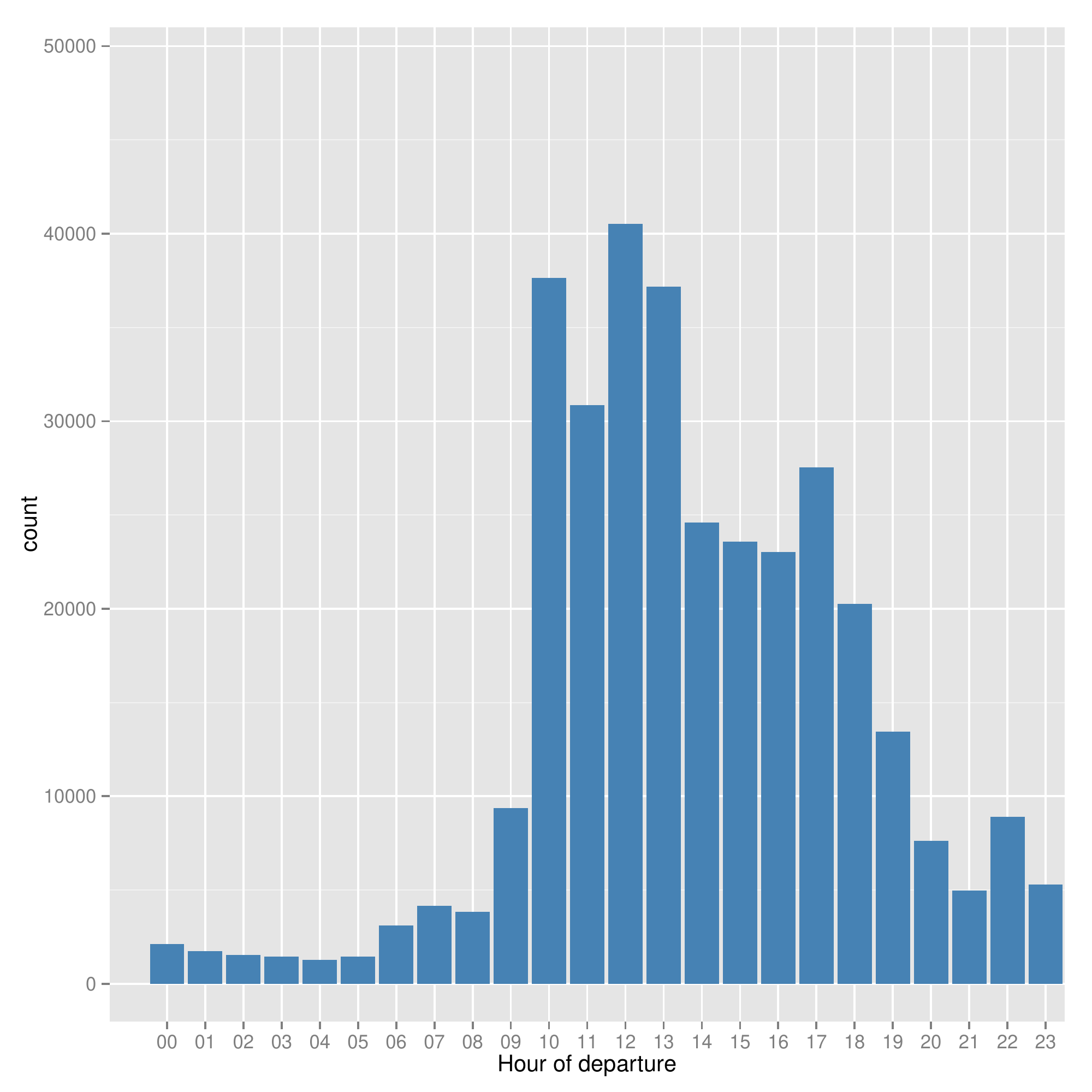}
\includegraphics[scale=0.23]{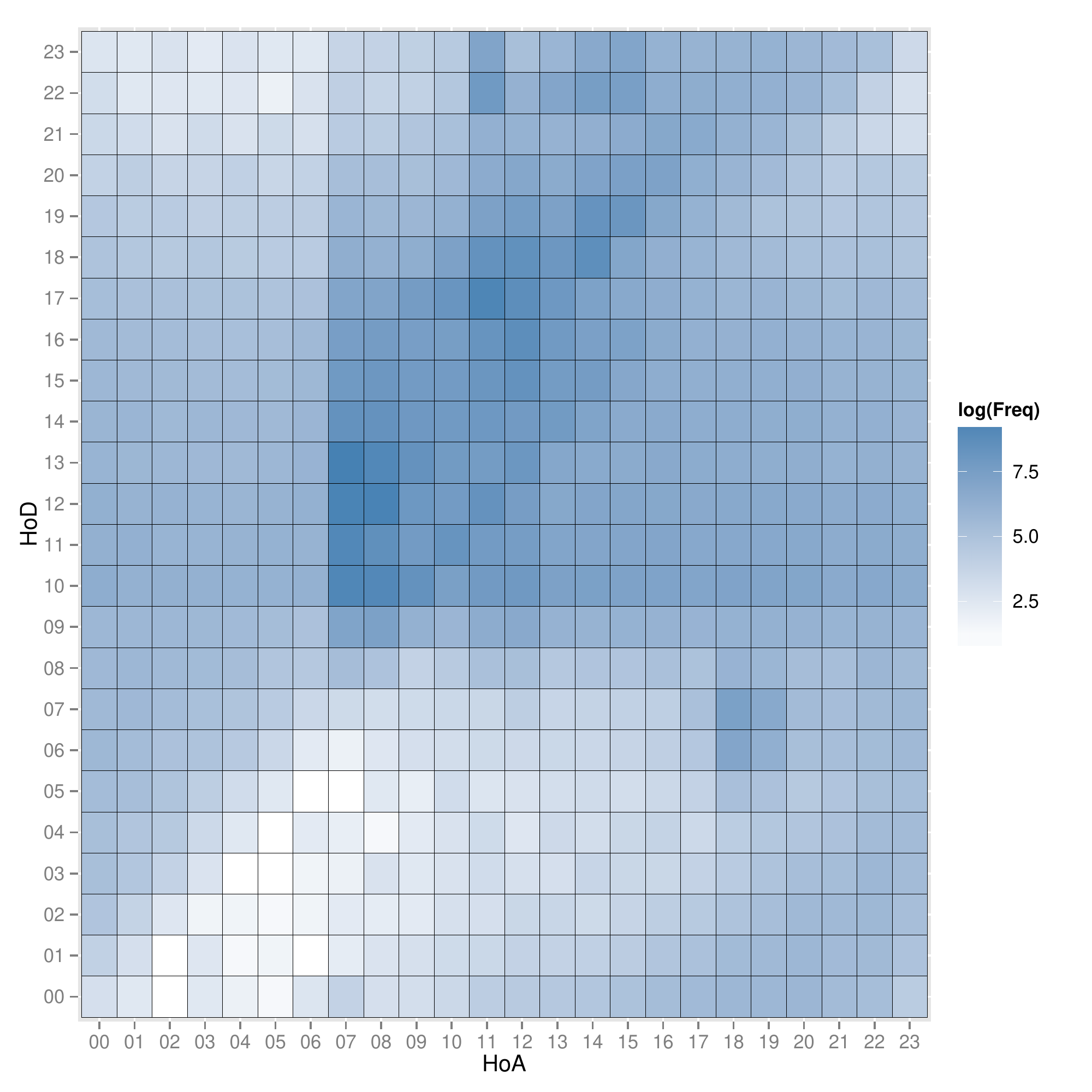}
\caption{\label{fig:hoahod} Comparison between time of arrival, time of discharge, and relationship between the two.}
\end{figure}

The objective of our work is to analyse LOS using a new approach, where the fitting and the medical interpretation of the results are the two aims. Our model is defined as a mixture of two distributions, one describing the short stays, and one the long stays. Moreover, the idea driving the "long-stay" component is to consider the observation as the results of the sum of two random variables, one describing the recovery period, and the other describing the discharge lag (linked to the hospital process). The "short-stay" component, on the other hand, will be fitted has a log-normal distribution as we assume that the discharge lag will be negligible. The results obtained in this work are of two kinds. First we prove that the fitting of the data obtained through this model is equivalent to the best existing fitting in the literature. Then, through regression results, we help identifying the features of influence on the discharge lag, hence pointing what is unusual in the hospital process. The paper is organized as follows. Section \ref{sec:model} contains a theoretical exposition of the modelling principles: the model description, and the importance of differentiating between the two effects. In Section \ref{sec:estimation} we present different estimation techniques. While the maximum likelihood remains the favourite technique, it can suffer from slowness. The Expectation-Maximisation algorithm, on the other hand, has good chances of reaching a local maximum if the initialization is not properly defined. We propose different options to sort this out. In Section \ref{sec:res}, we apply this method to real LOS data from a hospital in Victoria (Australia). Finally, we have included a short discussion on the proposed methodology.

\section{Modelling principles}
\label{sec:model}

The outcome variable LOS is usually defined to be the number of (whole) days from admission to discharge. This assumption of an integer value for the outcome variable limits the possible distributions that can be used to fit a model to the data. Moreover, we find this assumption too restrictive to understand what part of the LOS is due to the disease, and what part is due to the hospital process. By allowing the LOS variable to be continuous (with time unit still being the day), we stay closer to the hospital reality.\\ 
Let $\mathsf{Y}$ be the random variable representing the length of stay. We have $N$ observations, denoted $(\mathsf{y}_i)_{i \leq N}$. Each length of stay is linked to a hospital admission record, providing additional information that can help in understanding the random behaviour of $\mathsf{Y}$. Let denote these complementary observations $\mathsf{X}$.

\subsection{The short stay / long stay dichotomy}

The first modelling idea in this paper is to consider that the length of stay is following a mixture distribution, that is,
\begin{eqnarray}
\label{eq:mixture1}
\mathsf{Y} &=& \pi \mathsf{Y_L} + (1-\pi) \mathsf{Y_S}
\end{eqnarray} 
where $\pi$ is a random Bernoulli variable of unknown parameter value. $\mathsf{Y_L}$ (respectively $\mathsf{Y_S}$) is the variable associated with the long stays (resp. short stays). This model is quite straightforward as we can observe from Figure \ref{fig:hist} that there is a bump in the distribution of the small stays, and quite a heavy tail for long length of stays.\\
However, this model is not accurate enough as it fails to capture the variations in the long stays. To overcome this, we propose a second layer in our model by describing more accurately what we believe is the recovery process.

\subsection{The "long-stay" convolution model}

In our model, we assume that $\mathsf{Y_L}$ is the outcome of a two level process. The first component, the patient-disease related length of stay, will describe the recovery period part of the LOS. The second component, related to the hospital management, will affect the daily variations of LOS. As such, we can write $\mathsf{Y_L}$,
\begin{eqnarray}
\label{eq:model}
\mathsf{Y_L} &=& \mathsf{K} + \mathsf{E}
\end{eqnarray}
where $\mathsf{K}$ is a random variable on integer values (for example, a negative binomial distribution), and $\mathsf{E}$ is a continuous random variable (for example a Gaussian distribution). The regression principle is easily extensible to this case, and we have,
\begin{eqnarray}
\mathbb{E}[\mathsf{Y_L} | \mathsf{X}] &=& \mathbb{E}[\mathsf{K}| \mathsf{X}] + \mathbb{E}[\mathsf{E}| \mathsf{X}]
\end{eqnarray}
If we denote $f$ the density function, we then have,
 \begin{eqnarray}
 \label{eq:prob}
f_{\mathsf{Y_L}|\mathsf{X}}(\mathsf{y}) &=& \int f_{\mathsf{E}|\mathsf{X}}(\mathsf{y} - \mathsf{k}) f_{\mathsf{K}|\mathsf{X}}(\mathsf{k}) d\mathsf{k}
\end{eqnarray}
By assuming the conditional independence of the observations, the likelihood is simply defined as the product of the densities defined in Eq. \ref{eq:prob} incorporated into the mixture probability. A detailed expression of the likelihood is given in section \ref{ss:mle}.\\
As an example, using $K$ as a Negative Binomial and $E$ as a Gaussian, and replacing the densities with their expressions, we have,
\begin{eqnarray}
f_{\mathsf{Y_L}|\mathsf{X}}(\mathsf{y}) &=& \sum_{\mathsf{k} = 0}^{+\infty} \frac{1}{\sigma(\mathsf{x}) \sqrt(2 \pi)}\exp \big[ -\frac{1}{2 \sigma(\mathsf{x})^2}(\mathsf{y} -\mathsf{k} - m(\mathsf{x}))^2 \big] \frac{\Gamma(r(\mathsf{x}) + \mathsf{k})}{\Gamma(r(\mathsf{x})) \mathsf{k}!} p(\mathsf{x})^{r(\mathsf{x})} (1 - p(\mathsf{x}))^{\mathsf{k}}
\end{eqnarray}
As stated in the previous equation, the covariates $\mathsf{X}$ can be of influence in any of the for parameters of the convoluted distribution. Table \ref{tble:par} summarizes the modelling assumptions we are making by building a such model.
\begin{table}[t]
\small
\caption{\label{tble:par} Example of model description, with a Negative Binomial and a Normal distribution.}
\centering
\begin{tabular}{cp{4cm}lp{5.5cm}}
\toprule
Parameter & Parameter Interpretation & Ex. of variable & Variable Interpretation\\
\midrule
$p(\mathsf{x})$ & Probability of success & Admission day & On a weekday, more nurses and doctors can sign up on the discharge.\\ 
$r(\mathsf{x})$ & Number of failures before success & Care type & Depending on the care type, multiple exams may have to be made, hence an increased lag.\\
$m(\mathsf{x})$ & Average recovery period & DRG & A patient with a particular disease type is expected to recover in $m$ days.\\
$\sigma(\mathsf{x})$ & Variation of recovery period & Patient age & Older patients may have more variation in their recovery period.\\
\bottomrule
\end{tabular}
\end{table}
This formulation allows complex functions to be fitted, also we will focus in this paper on linear functions, 
\begin{eqnarray}
s(\mathsf{x}) & = & h(\beta_{s} \mathsf{X}_{s})
\end{eqnarray}
where $s = \{p,r,m,\sigma\}$ and $h$ will serve as a transform to ensure the linear combination does not produce outbound parameters.\\
The combination of possible distributions is described below. For the recovery period, we can use:
\begin{itemize}
\item The normal or the log-normal distribution, described by their parameters $\mu$ and $\sigma^2$, and which densities are
\begin{equation}
f_1({\sf x}) = \frac{1}{\sigma \sqrt{2 \pi}} \exp\Big[\frac{-({\sf x} - \mu)^2}{2\sigma^2} \Big] \qquad \textrm{and} \qquad f_2({\sf x}) = \frac{1}{ {\sf x}\sigma \sqrt{2 \pi}} \exp\Big[\frac{-(\ln {\sf x} - \mu)^2}{2\sigma^2} \Big]
\end{equation}
\end{itemize}
For the discharge lag, we can use:
\begin{itemize}
\item The Negative Binomial distribution, described by its parameters $r$ and $p$, and which density is
\begin{equation}
p_1({\sf k}) = \frac{\Gamma(r+{\sf k})}{\Gamma(r) \Gamma({\sf k})} p^r (1-p)^{\sf k} 
\end{equation}
This distribution is helpful to model over-dispersion of counts data.
\item The Poisson distribution, described by its parameter $\lambda$, and which density is
\begin{equation}
p_2({\sf k}) = \frac{\lambda^{\sf k}}{{\sf k}!} \exp[-\lambda] 
\end{equation}
The Poisson distribution is the most used distribution to model counts.
\item The Conway Maxwell Poisson distribution, described by its parameters $\lambda$ and $\nu$, and which density is
\begin{equation}
p_3({\sf k}) = \frac{\lambda^{\sf k}}{({\sf k}!)^{\nu}} \frac{1}{Z(\lambda, \nu)}
\end{equation}
where $Z()$ is a normalizing constant. This distribution, well less foreknown than the two previous ones, is useful to model under-dispersion in counts data.
\end{itemize}
Then, we can also use distribution with bounded support, making the approach more like a classical mixture problem,
\begin{itemize}
\item The Binomial distribution, described by its parameters $n$ and $p$, and which density is
\begin{equation}
p_4({\sf k}) = C^{\sf k}_n p^{\sf k} (1-p)^{n - {\sf k}}
\end{equation}
\item The Multinomial distribution, described by its parameters $p_1$ to $p_K$, where $K+1$ is the maximum number of groups. Its density is
\begin{equation}
p_5({\sf k}) = p_{\sf k}
\end{equation}
\end{itemize}

\subsection{Link with image processing}
\label{mp:image}
This convolution-based approach is nothing new in the signal processing literature, in particular for the image processing scientists. In their context, the convolution model aims at describing the degradation of an image due to noise. One particular model aims at describing the degradation in two components. The Poisson component, which relates to the quantum nature of lights, models the impulsive noise, and the Gaussian component, which relates to the noise present in the electronic part of the imaging system \citep[see][]{Jezierska2013, Xiao2011, Yan2013}. The aim of these approaches is to recover the image and estimate the noise parameters altogether. Different algorithms have been developed to solve this inverse problem, which is often described by the following equation,
\begin{eqnarray}
\min_{\mathsf{x} \in \mathbb{R}^N} \big( f(\mathsf{x}) = \Phi(\mathsf{x}) + \rho(\mathsf{x}) \big)
\end{eqnarray}
where $\Phi$ stands for the data fidelity term, and $\rho$ is a regularization function incorporating a priori information. In the Bayesian framework, this allows to compute the MAP estimate of the original signal $\mathsf{x}$. In this context, the data fidelity term is defined as the negative logarithm of the likelihood and the regularization term corresponds to the potential of the chosen prior probability density function.\\
In the context of this article, the original signal can be viewed as the disease-related length of stay before its degradation by the hospital process. Therefore, an estimation of the disease-related length of stay would not only give an idea of the hospital stress (by differentiating with the actual LoS) but could also be used as a first-hand classification tool for the disease. Not to mention the hospital management evolutions that can be derived from such an information.

\section{Estimation procedure}
\label{sec:estimation}
The study has a principal aim. We want to forecast the LOS of a patient arriving in a hospital according to the information available upon arrival. In order to refine the prediction, we need to correctly identify the covariates of importance, and infer their influence. This will be done through the parameter estimation (maximum likelihood estimation, in section \ref{ss:mle} and expectation-maximisation algorithm, section \ref{ss:em}), which will the allow for prediction. The model developed in this paper makes it also possible to identify the possible hospital stress. This is made possible by the separation of the disease-patient related LOS from the delays due to the hospital process. Therefore, by carefully analysing the estimated parameters of the hospital-linked features, we may be able to identify the stress factors in the LOS, if existing.

\subsection{Likelihood based estimation}
\label{ss:mle}
With the notations from the previous section, and assuming that the LOS observations are conditionally independent from each other, we can write the log-likelihood,
\begin{eqnarray}
\label{eq:logl}
\log \mathsf{L}(\mathsf{y}_{i=1\dots N} | {\bm \theta}) &=& \sum_i \log f_{{\bm \theta}}(\mathsf{y}_i) \nonumber\\
{} &=& \sum_i \log \Big[p f_{\mathsf{Y_S}|\mathsf{X}}(\mathsf{y}_i) + (1-p) f_{\mathsf{Y_L}|\mathsf{X}}(\mathsf{y}_i)\Big]\nonumber\\
{} &=&\sum_i \log \Big[p f_{\mathsf{Y_S}|\mathsf{X}}(\mathsf{y}_i) + (1-p) \sum_\mathsf{k} f_{\mathsf{E}|\mathsf{X}}(\mathsf{y} - \mathsf{k}) f_{\mathsf{K}|\mathsf{X}}(\mathsf{k}) \Big]
\end{eqnarray}


Maximizing Eq. \ref{eq:logl} is a very hard task, that has to be handled numerically \cite{Simar1976, Sprott1983}. But even numerically, the convergence of the existing algorithms can be quite slow, in particular when the dimension of $\mathsf{X}$ increases. Wager \cite{Wager2013} proposes a geometrical approach for estimating the parameters, using the empirical distribution, which roughly matches the accuracy of fully general maximum likelihood estimators at a fraction of the computational cost.

\subsection{Expectation-maximisation (EM) algorithm}
\label{ss:em}
The baseline model we use being a mixture, a natural estimation procedure would be the expectation-maximization (EM) algorithm. The EM algorithm\cite{Dempster1977} is the most popular approach for calculating the maximum likelihood estimator of latent variable models. For a thorough review on the estimation of mixture models, see \cite{Redner1984, Bilmes1998, Marin2005}. However, due to the nonconcavity of the likelihood function of latent variable models, the EM algorithm generally only converges to a local maximum rather than the global one \cite{Wu1983}. On the other hand, existing statistical guarantees for latent variable models are only established for global optima \cite{Bartholomew2011}.\\
Let briefly review the classical EM algorithm. Given the observations $(\mathsf{y}_i)_{i \leq N}$, and an unobserved latent variable $\mathsf{Z} \in \mathcal{Z}$, the algorithm aims at maximizing the augmented log-likelihood
\begin{eqnarray}
\ell_N({\bm \theta})= \sum_i \log h_{\bm \theta}(\mathsf{y}_i) & \textrm{where} & 
h_{\bm \theta}(\mathsf{y}) = \int_{\mathcal{Z}} f_{\bm \theta} (\mathsf{y}, \mathsf{z}) d\mathsf{z} 
\end{eqnarray}
Because $\mathsf{Z}$ is unobserved, it is difficult to evaluate $\ell_N({\bm \theta})$. Instead, the EM maximize
\begin{eqnarray}
\bm{Q}_N({\bm \theta}; {\bm \theta'}) &=& \frac{1}{N} \sum_i \int_\mathcal{Z} f_{\bm \theta}(\mathsf{z} \vert \mathsf{y_i}) \log f_{\bm \theta}(\mathsf{y}_i, \mathsf{z}) d\mathsf{z}
\end{eqnarray}
which is the difference between the optimal likelihood and a given likelihood. In order to achieve the maximisation, the EM proceeds in two steps during each iteration,
\begin{equation}
\left|
\begin{array}{lcl}
\textrm{\bf(E)} & \textrm{Compute} & f_{\bm \theta}(\mathsf{z} \vert \mathsf{y_i}) \\
\textrm{\bf (M)} & \textrm{Set} & {\bm \theta}^{(c+1)} = \underset{\bm \theta}{\operatorname{argmax}} \quad \bm{Q}_N({\bm \theta}, {\bm \theta}^{(c)})
\end{array}
\right.
\end{equation}
In the context of this article, $\mathsf{Z}$ will describe the short or long stays, meaning $\mathcal{Z} = \{0,1\}$. With the previous notations, we would also have,
\begin{eqnarray}
\label{eq:em2}
f_{\bm \theta}(\mathsf{z} = 0 \vert \mathsf{y_i})\propto p(\mathsf{z} = 0) \times f_{\mathsf{Y_S}|\mathsf{X}}(\mathsf{y}_i)  & \textrm{and} & f_{\bm \theta}(\mathsf{z} = 1 \vert \mathsf{y_i})\propto  p(\mathsf{z} = 1)\times \sum_\mathsf{k} f_{\mathsf{E}|\mathsf{X}}(\mathsf{y} - \mathsf{k}) f_{\mathsf{K}|\mathsf{X}}(\mathsf{k})
\end{eqnarray}
which will be normalised to keep the right properties, and
\begin{eqnarray}
f_{\bm \theta}(\mathsf{y_i},\mathsf{z}) &=& \int_{\mathcal{Z}} f_{\bm \theta}(\mathsf{y}_i \vert \mathsf{z}) p(\mathsf{z}) d\mathsf{z}
\end{eqnarray}
\subsection{Two-dimensional Expectation-maximisation (2d-EM) algorithm}
\label{ss:2d-em}

A different interpretation of the model can also lead to the idea of a double EM algorithm. Indeed, the model stated in eq. \ref{eq:mixture1} decribes the mixture of two distributions, one of them being also a mixture (while infinite, it is a mixture, sometimes also called convolution). In this double mixture model framework, we increase the dimension of the latent variable, so that $\mathsf{Z} = (\mathsf{C}, \mathsf{S}) \in \mathcal{C} \times \mathcal{S}$. If $\mathsf{C}$ and $\mathsf{S}$ were independent, the problem would we the same as a normal latent variable problem. However, two differences exist here. First, $\mathsf{C}$ and $\mathsf{S}$ are not independent. Also, $\mathcal{C} = \mathbb{N}$, making one of the mixture infinite. The main driver behind that interpretation of the model is the convolution distribution in the following equation,
\begin{eqnarray}
\label{eq:em3}
\sum_\mathsf{k} f_{\mathsf{E}|\mathsf{X}}(\mathsf{y} - \mathsf{k}) f_{\mathsf{K}|\mathsf{X}}(\mathsf{k})
\end{eqnarray}
This probability density function can be long to calculate, in particular when $\mathsf{k} \in \mathbb{N}$. To avoid this additional computational complexity, we will use this double EM to use only analytical density functions.\\
The literature on infinite mixtures finds solution in the Bayesian community working on Gaussian processes \cite{Rasmussen1999, Rasmussen2002}. While a maximum likelihood estimation is quite common in finite mixture problems \cite{Zhang2009}, its generalization to infinite mixture is not straightforward. Using Gaussian and Dirichlet processes, Sun \cite{Sun2011} proposes different inference techniques to fit the model. In the end, all the proposed methods refers to the EM algorithm as a way to deal with the incomplete nature of the data. Moulines \cite{Moulines1997} briefly mentions the link between deconvolution and mixture, proposing an EM algorithm to estimate both the noise and the signal parameters.  The link with the mixtures is quite tenuous because of the infinite nature of the sum in the likelihood equation. Moreover, the usual mixture definition implies an empirical estimation of the mixture probabilities, which make the infinite assumption impossible to estimate. However, with a parametric assumption on the mixture weights, it becomes theoretically and practically possible.\\
Following the change in the latent variable dimension, we can rewrite $\bm Q$,
\begin{eqnarray}
\label{eq:Q}
\bm{Q}_N({\bm \theta}; {\bm \theta'}) &=& \frac{1}{N} \sum_i \int_\mathcal{C} \int_\mathcal{S}  f_{\bm \theta}(\mathsf{c},\mathsf{s} \vert \mathsf{y_i}) \log f_{\bm \theta}(\mathsf{y}_i, \mathsf{c}, \mathsf{s}) d\mathsf{c} d \mathsf{s}
\end{eqnarray}
In this expression, we still need to calculate the two terms. First, we recall that we can re-write
\begin{eqnarray}
 f_{\bm \theta}(\mathsf{y}_i, \mathsf{c}, \mathsf{s}) &=& f_{\bm \theta}(\mathsf{y}_i \vert \mathsf{c}, \mathsf{s}) \times f_{\bm \theta}(\mathsf{c} \vert \mathsf{s})  \times f_{\bm \theta}(\mathsf{s}) 
\end{eqnarray}
$f_{\bm \theta}(\mathsf{s})$ is the mixing parameter (defining the absolute probability of $\mathsf{s}$) that will be non-parametrically estimated using the standard EM procedure for the mixing coefficients. Then we calculate $f_{\bm \theta}(\mathsf{c} \vert \mathsf{s})$ such that,
\begin{equation}
\label{eq:Q2}
\forall \mathsf{c} \in \mathcal{C}, \left\{
\begin{array}{lll}
f_{\bm \theta}(\mathsf{c} \vert \mathsf{s} = 0) & = & \delta_1(\mathsf{c})\\ 
f_{\bm \theta}(\mathsf{c} \vert \mathsf{s} = 1) & = & f_{\mathsf{K} \vert \mathsf{X}}(\mathsf{c})
\end{array}
\right.
\end{equation}
Identically, we define $f_{\bm \theta}(\mathsf{y}_i \vert \mathsf{c}, \mathsf{s})$,
\begin{equation}
\label{eq:Q3}
\forall \mathsf{c} \in \mathcal{C}, \left\{
\begin{array}{lll}
f_{\bm \theta}(\mathsf{y}_i \vert \mathsf{c}, \mathsf{s}=0) & = & f_{\mathsf{Y_S}|\mathsf{X}}(\mathsf{y}_i)\\ 
f_{\bm \theta}(\mathsf{y}_i \vert \mathsf{c}, \mathsf{s}=1) & = & f_{\mathsf{E}|\mathsf{X}}(\mathsf{y}_i - \mathsf{c})
\end{array}
\right.
\end{equation}
Finally, we have to calculate $f_{\bm \theta}(\mathsf{c},\mathsf{s} \vert \mathsf{y_i})$. Because $\mathsf{s}$ and $\mathsf{c}$ are not independent, we use once again the conditioning (and the Bayes paradigm), yielding,
\begin{eqnarray}
f_{\bm \theta}(\mathsf{c},\mathsf{s} \vert \mathsf{y_i}) &\propto & f_{\bm \theta}(\mathsf{y}_i \vert \mathsf{c}, \mathsf{s}) \times f_{\bm \theta}(\mathsf{c} \vert \mathsf{s})  \times f_{\bm \theta}(\mathsf{s}) 
\end{eqnarray}
Thanks to Eqs. \ref{eq:Q2} and \ref{eq:Q3}, we can achieve our algorithm. We can now define the following three steps of the algorithm,
\begin{equation}
\left|
\begin{array}{lcl}
\textrm{\bf(E-1)} & \textrm{Compute} & f_{\bm \theta}(\mathsf{s} \vert \mathsf{y_i}) \\
\textrm{\bf(E-2)} & \textrm{Compute} & f_{\bm \theta}(\mathsf{c} \vert \mathsf{y_i}, \mathsf{c}) \\
\textrm{\bf (M)} & \textrm{Set} & {\bm \theta}^{(c+1)} = \underset{\bm \theta}{\operatorname{argmax}} \quad \bm{Q}_N({\bm \theta}, {\bm \theta}^{(c)})
\end{array}
\right.
\end{equation}

\section{Results}

\subsection{Fitting: Comparison with previous studies}
\label{sec:prev}

\begin{figure}[t]
\centering
\begin{subfigure}[b]{0.45\textwidth}
\includegraphics[scale=0.20]{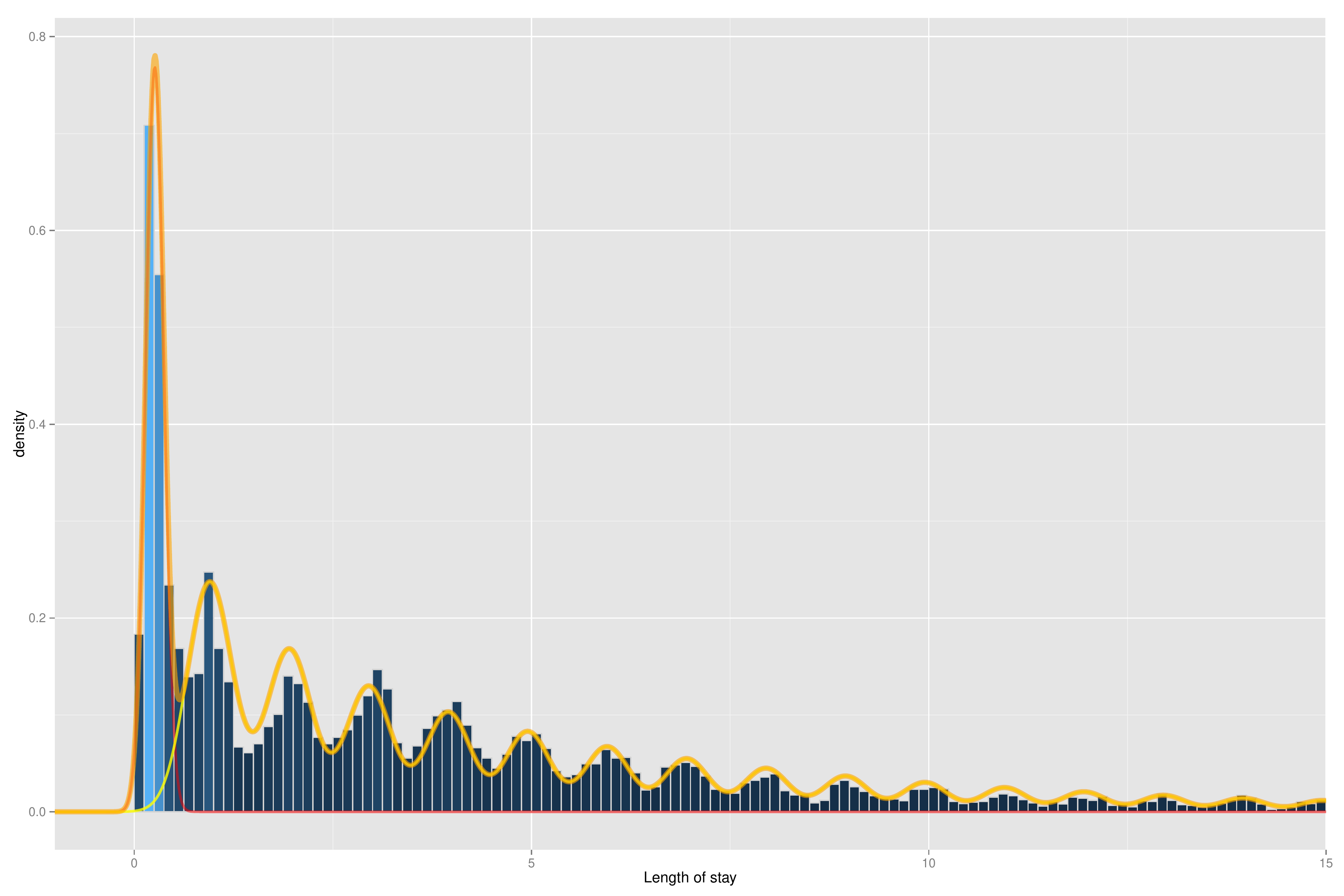}
\caption{Full data fitting}
\end{subfigure}
\begin{subfigure}[b]{0.45\textwidth}
\includegraphics[scale=0.20]{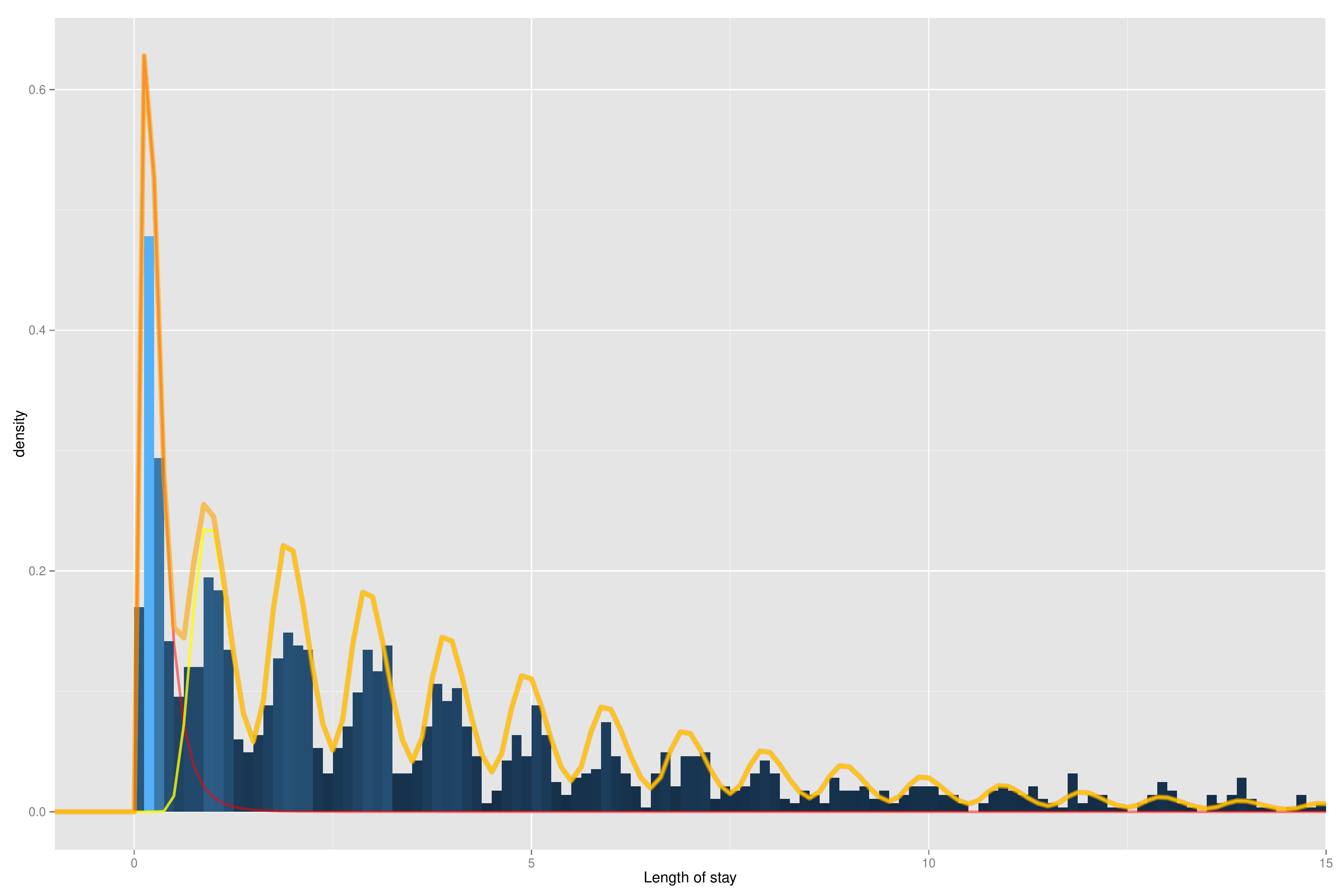}
\caption{DRG $14$}
\end{subfigure}\\
\begin{subfigure}[b]{0.45\textwidth}
\includegraphics[scale=0.20]{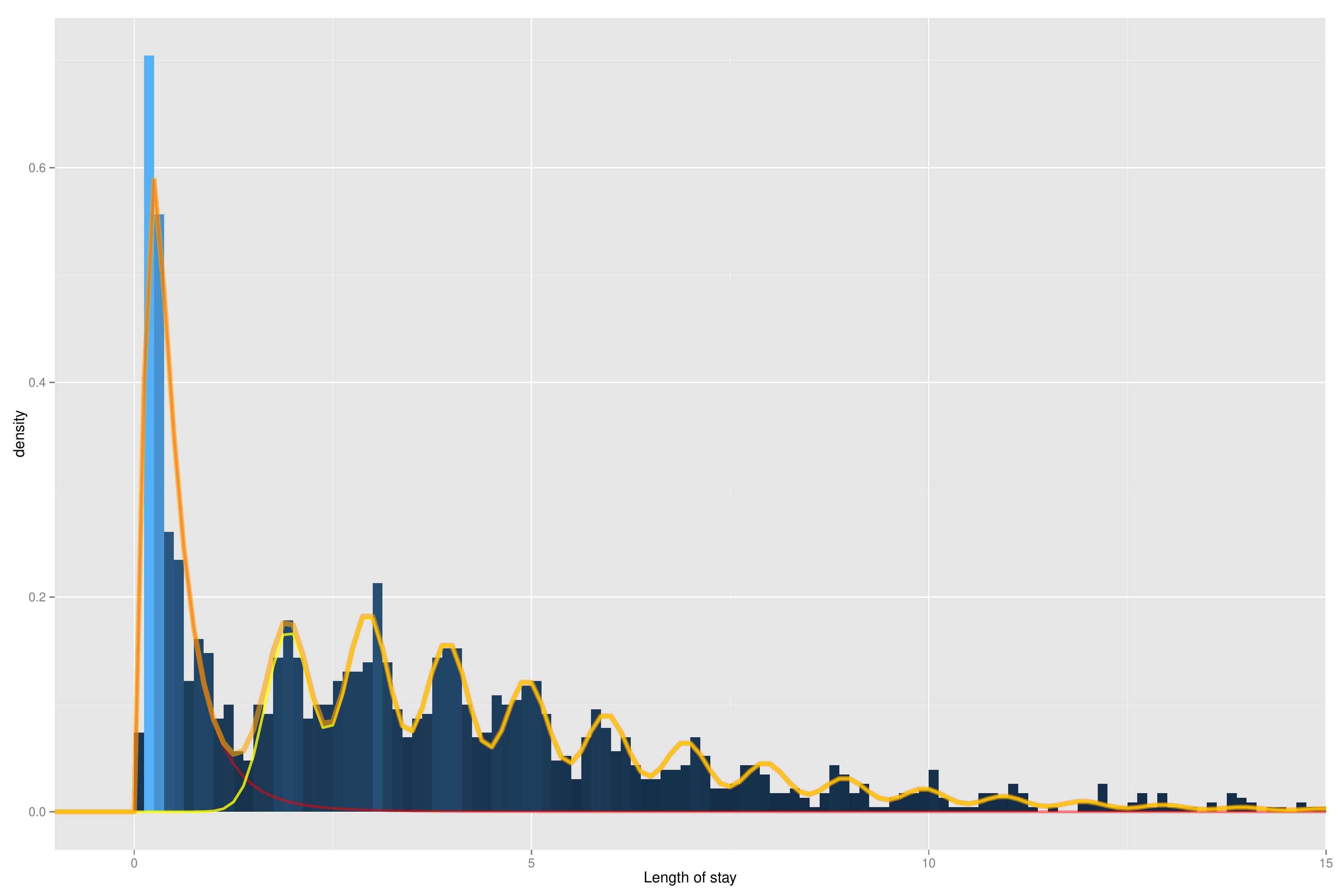}
\caption{DRG $88$}
\end{subfigure}
\begin{subfigure}[b]{0.45\textwidth}
\includegraphics[scale=0.20]{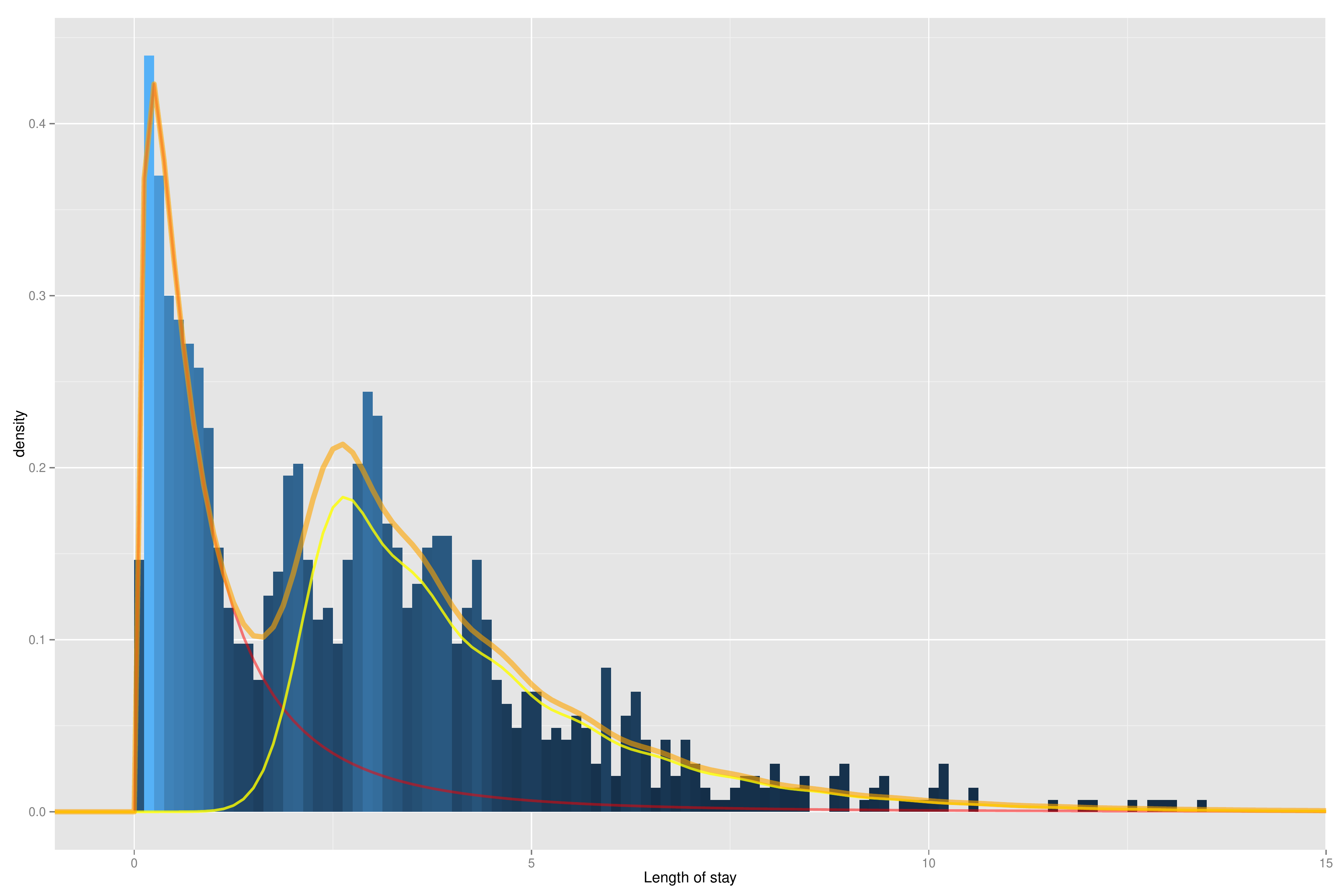}
\caption{DRG $122$}
\end{subfigure}\\
\begin{subfigure}[b]{0.45\textwidth}
\includegraphics[scale=0.20]{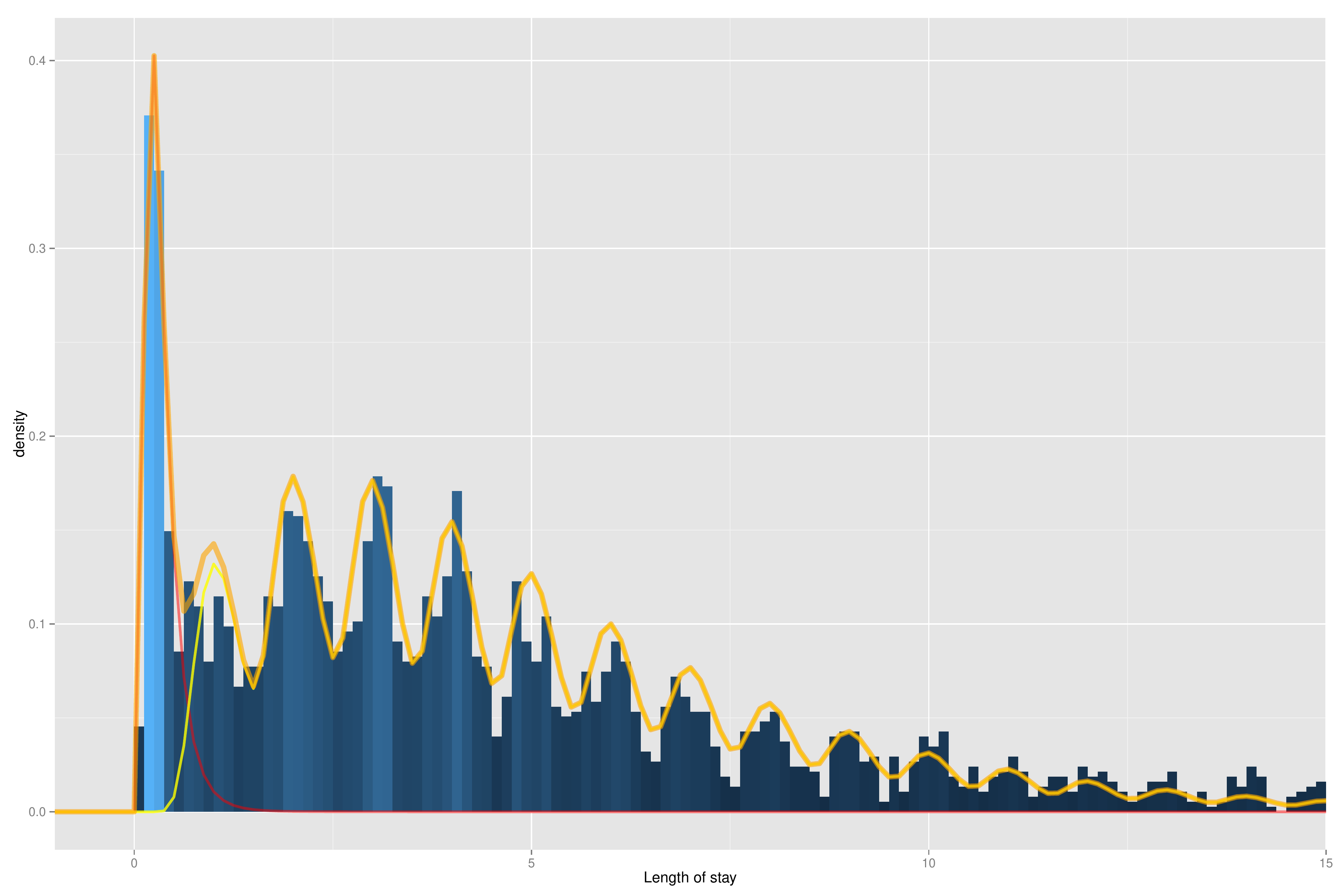}
\caption{DRG $127$}
\end{subfigure}
\begin{subfigure}[b]{0.45\textwidth}
\includegraphics[scale=0.20]{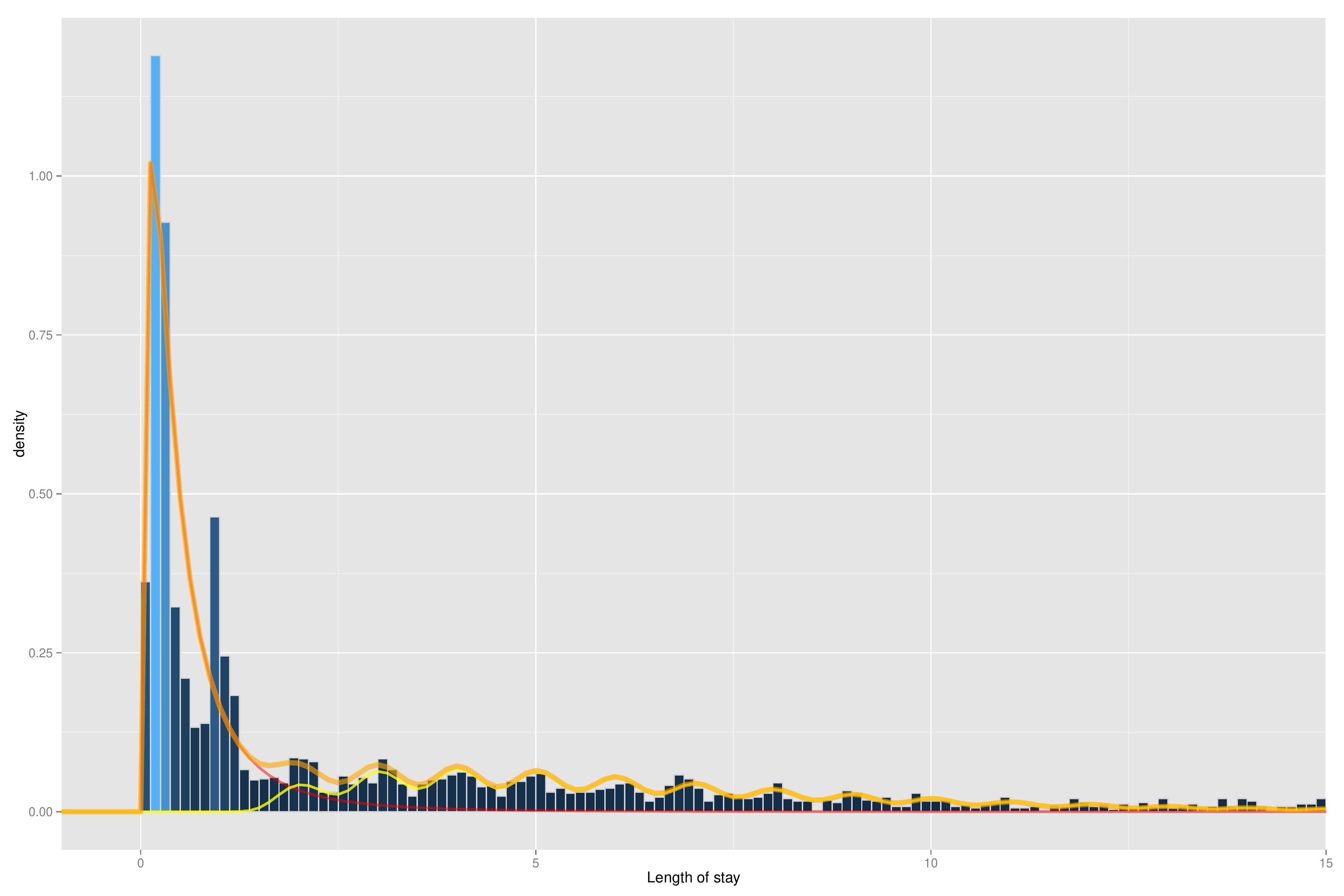}
\caption{DRG $541$}
\end{subfigure}
\caption{\label{fig:disc} Graphics on the modelization of LOS variable for several DRGs. (b)-(f): fitting of the estimated distribution and histogram of observations for each individual DRG. (a) Fitting for all the data. The yellow lines represent the long stay distributions, and the red lines the short stay distributions.}
\end{figure}

In \citep{Atienza2008} the authors fit a (finite) mixture of different distributions (Gamma, Weibull, Lognormal) to the data from different DRGs. In order to compare our model to theirs, we will use our data issued from the same DRGs:
\begin{itemize}
\item DRG $14$: Specific cerebrovascular disorders except transient ischemic
attack (DRGs $B70$ in our data);
\item DRG $88$: Chronic obstructive pulmonary disease (DRG $E65B$ in our data);
\item DRG $122$: Circulatory disorders with acute myocardial infarction without cardiovascular complications, discharged alive (DRG $F60B$ in our data);
\item DRG $127$: Heart failure and shock (DRGs $F62A$ and $F62B$ in our data);
\item DRG $541$: Respiratory diseases except infection, bronchitis and asthma (DRGs $E02$, $E40$, $E41Z$, $E64$, $E67$, $E71$, $E75$, $E76Z$ in our data),
\end{itemize}
and compute a discrepancy measure. As a discrepancy measure between the empirical $F_{\theta_e}(.)$ and the estimated distribution function $F_{\hat{\theta}}(.)$, the uniform measure \citep{Zolotarev1983}, also called Kolmogorov measure, is proposed:
\begin{eqnarray}
d(\theta_e, \hat{\theta}) &=& \sup_{x \in \mathbb{R}} \vert F_{\theta_e}(x) - F_{\hat{\theta}}(x) \vert
\end{eqnarray}
The results are presented in Table \ref{tble:discrepancy}, for different distributions and the proposed DRGs.
\begin{table}[t]
\small
\caption{\label{tble:discrepancy} Measure of distance between the estimated distribution and the empirical distribution.}
\centering
\begin{tabular}{lcccccc}
\toprule
Model & All $5$ DRGs & DRG $14$ & DRG $88$ & DRG $122$ & DRG $127$ & DRG $541$\\
\midrule
Lognormal & 0.096 & 0.045 & 0.112 & 0.111& 0.086 & 0.087\\
Gamma & 0.064 & 0.107 & 0.076 & 0.123 & 0.114 & 0.139\\
Weibull & 0.063 & 0.077 & 0.063 & 0.092 & 0.077 & 0.091\\
Mixture \citep{Atienza2008} & 0.023 & 0.018 & 0.049 & 0.108 & {\bf 0.017} & 0.054\\
Proposed model & {\bf 0.019} & {\bf 0.011} & {\bf 0.016} & {\bf 0.023} & 0.019 & {\bf 0.022}\\
\bottomrule
\end{tabular}
\end{table}
Amongst the fitted distributions, we choose the most popular choices, namely the Log-normal, the Gamma and the Weibull distribution. Moreover, we compare our performance with the best existing in the literature, the model proposed in \citep{Atienza2008}. The proposed model is the chosen to be the best fit amongst the combination of Poisson-Negative Binomial-CoMPoisson and Gaussian-Log Normal. It can also be chosen amongst the combination of Binomial and Multinomial distributions. The main difference between these models is the nature of the discrete distribution (and hence the mixture), which has an infinite support in the first case and a finite support in the later. The choice of the model has a significant influence and has to be made with a purpose in mind. For a better fitting and understand, the Multinomial approach can prove very useful. However, if the aim is the forecast, the infinite mixture may prove more robust, in particular in situations where few data are available. As we can see from the table, the results of our model are outperforming the results of the model proposed in \citep{Atienza2008}, except the distribution in DRG $127$ where the results can be considered equivalent. We also display in Figure \ref{fig:disc} the density histograms for the $5$ DRGs (subplots (a) for all DRGs, (b)-(f) for each individual DRG), with curves representing the estimated distributions. 

\subsection{Understanding: Explaining LOS variations}
\label{sec:res}

\subsubsection*{List of features}

The interpretation of our statistical modelling is to consider that a particular disease on a particular patient will need to be treated in a number of day that will be distributed as the recovery period. Then, additional noise is considered, due to the hospital processes, that will be distributed according to a distribution whose support will be finite or infinite. To fit that model, we have the features presented in Table \ref{tble:feat}.\\

\begin{table}[t]
\small
    \caption{\label{tble:feat}Table of possible explanatory variable for LoS.}
    \begin{subtable}{.5\linewidth}
      \centering
        \caption{Patient related features}
        \begin{tabular}{llc}
\toprule
Feature name & feature type & dimension\\
\midrule
Age & numerical & .\\
Gender & categorical & $3$ \\ 
Marital Status & categorical & $4$ \\
Ethnicity & categorical & $5$ \\
Country & categorical & $50$ \\
Disease Type & categorical & $9$ \\
DRG & categorical & $634$\\
\bottomrule
\end{tabular}
    \end{subtable}%
    \begin{subtable}{.5\linewidth}
      \centering
        \caption{Hospital related features}
        \begin{tabular}{llc}
\toprule
Feature name & feature type & dimension\\
\midrule
Hour of arrival & numerical & . \\
Day of arrival & categorical & $7$\\
Month of arrival & numerical & . \\
Admission Type & categorical & $4$ \\
Admission Unit & categorical & $46$ \\
Discharge Unit & categorical & $46$ \\
Care Type & categorical & $4$ \\
\bottomrule
\end{tabular}
    \end{subtable} 
\end{table}

We also have made a number of assumption that will limit the number of parameters to be estimated. This is safer, as the estimation procedure can take a long time and some features have some really prohibitive dimensions. This list is:
\begin{itemize}
\item[-] The probability that a stay is a short or long stay cannot be explained by the available feature. This assumption is probably the weakest assumption (in terms of modelling) as it seems obvious that DRG, or Admission Unit will be important. This is clearly an assumption that will be removed in further work.
\item[-] The short stay parameters (mean and variance) will not depend on any of the available features. This modelling assumption is based on the fact that the very small variance of that distribution makes it useless to fit features.
\item[-] For the same reasons, the variance of the recovery period of the long stays component will not depend on any of the available features either.
\end{itemize}
For many reasons, the DRG cannot be used as is into the model. To overcome that problem, the data can be analysed by DRG, as we did in the previous section, or, more safely, by considering an intermediate kind of information. For example, the use of the first two letters of the DRG code can prove useful and not too vast.

\subsubsection*{Results and discussion}

The results of the model for the $5$ DRGs altogether are presented in Table \ref{tble:modelIestimation}. We observe in particular the specifics for the $5$ selected DRGs in Table \ref{tble:resDRGs}
\begin{table}[t]
\small
\caption{\label{tble:resDRGs} Numerical results of the model, with the repartition between short and long stayers, and the properties of these stays.}
\centering
\begin{tabular}{lccc|ccc}
\toprule
 & \multicolumn{3}{c}{Short stayers} &  \multicolumn{3}{c}{Long stayers}\\
 & $\%$ & mean & sd & $\%$ & mean & sd \\
\midrule
DRG $14$ & 20.4 & 6 hours & 3 hours & 79.6 & 4 days, 10 hours & 3 days, 10 hours\\
DRG $88$ & 24.8 & 7 hours & 3 hours & 75.2 & 4 days, 11 hours & 2 days, 20 hours\\
DRG $122$ & 45.6 & 17 hours & 11 hours & 54.4 & 4 days, 4 hours & 2 days\\
DRG $127$ & 14.9 & 7 hours & 3.5 hours & 85.1 & 4 days, 23 hours & 3 days, 6 hours\\
DRG $541$ & 63.0 & 13 hours & 11 hours & 37.0 & 6 days, 14 hours & 3 days, 14 hours\\
\bottomrule
\end{tabular}
\end{table}

\begin{landscape}
\begin{table}[t]
\caption{\label{tble:modelIestimation} Results of the regression model for the $5$ selected DRGs. The four results presented provide the influence of the predictors on (a) the probability of having a short stay, (b) the short stay duration, (c) the recovery period duration (long stay component) (d) the discharge lag (long stay component).}
\begin{subtable}{.45\linewidth}
      \centering
        \caption{Logistic regression results}
        \pgfplotstableread{coefficients_logit_all.txt}\data
        \input{table1}
    \end{subtable}%
    \vspace{0.1cm}
    \begin{subtable}{.45\linewidth}
      \centering
        \caption{Log-normal regression results}
        \pgfplotstableread{coefficients_logn_all.txt}\data
        \input{table1}
    \end{subtable} 
    \vspace{0.1cm}
    \begin{subtable}{.45\linewidth}
      \centering
        \caption{Recovery period regression}
        \pgfplotstableread{coefficients_recov_all.txt}\data
        \input{table1}
    \end{subtable}
    \hfill
    \begin{subtable}{.5\linewidth}
      \centering
        \caption{Discharge lag regression}
        \pgfplotstableread{coefficients_disch_all.txt}\data
        \input{table1}
    \end{subtable} 
\end{table}
\end{landscape}

\section{Discussion}

We presented in this article a new model of hospital length of stay data. This new model provides a better fitting of the data, and also gives a realistic description of the process producing the length of stay. We strongly believe that this model will prove useful for clinicians and hospital managers in their attempt to improve the patients and the medical staff experience.\\
This model is complex to estimate, due to the mixture and the convolution. The maximum likelihood approach is proposed, and is able to estimate the parameters correctly. However, it cannot tell us if the patient's stay belong to the short or the long category. This kind of information would typically be useful to identify outliers, or to perform additional statistical analysis on one or the other category. For this reason, we used the classical EM algorithm \cite{Dempster1977}. We faced another challenge with this estimation procedure, because the convolution distribution optimal solution can only be calculated numerically, which leads to extended delays in the optimisation procedure. To overcome this, we considered the convolution as an infinite mixture, where the mixture coefficient are parametrically defined. To perform the estimation, we proposed an augmented EM algorithm called 2d-EM. In this algorithm, we consider that the dimension of the latent space is equal to $2$. This allows more flexibility in the model, while the computational complexity remains manageable. Finally, we applied it to real data from Melbourne (Australia). We identify important variables for the purpose of length of stay modelling. 
\\
Another aspect that must be discussed is that the proposed model is a complicated one. As research scientists working with real world problems, we have to make a decision between the complexity of the model / computation procedure and the fitting of the data / meaningfulness of the model. For this model, the computational effort and the complexity should not be considered as excessive. Furthermore, this methodology seems more adequate for the DRGs we have worked on. The model based on a unique family of distributions is less complex but has two drawbacks. First, the necessity of previously selecting the family among the most usual asymmetric distributions. Second, the results provided are less optimal. Therefore, the study of a simpler model does not imply either a significant reduction of the computational effort or better results.\\
A few more words on the understanding of the proposed model are needed. First, regarding the optimality of the solution obtained by maximising the likelihood. At best, we have four parameters that need to be fitted. In some particular situation, that may lead to an overfitting of the data. We recommend the user to pay attention to the results, and any prior knowledge about the data should be considered with care. For example, the variance of the second distribution of the long stay model (usually a Gaussian distribution) can be specified, so that the estimation procedure ends quicker with an appropriate fit. Or at least, specify bounds in which that parameter should belong. The same careful consideration must lead the choice of the features and which variable ($\mathsf{K}$ or $\mathsf{E}$) they should be fitted in. Not only the estimation results will be impacted, but also their interpretations.\\
~\\
In conclusion, we believe that this work contributes to the development of the statistical analysis of LOS distributions and other consumption variables in health services. Also this approach can be applied to other asymmetric data (for instance, the length of wait for surgical procedures or for medical attention).

\section*{Acknowledgement}

The authors would like to acknowledge the financial support of Commonwealth Health Department for the project and in-kind operational support from Austin Health and the Royal Melbourne Hospital.

\bibliographystyle{wileyj}
\bibliography{/home/ick003/Documents/Donnees/References/Publications-LengthOfStay.bib}

\end{document}

%% file: mytikz.tex
\usepackage{tikz}
\usepackage{tikz-3dplot}
\usetikzlibrary{shapes.geometric, calc}
\usetikzlibrary{matrix}
\usetikzlibrary{trees}
\usetikzlibrary{chains}
\usetikzlibrary{positioning}
    \everymath{\displaystyle}
    \tikzstyle{mathbox} = [inner sep=0pt, anchor=base]
    \tikzstyle{every picture}+=[remember picture]
    \usetikzlibrary{calc}
    \usetikzlibrary{positioning}
    \xdefinecolor{darkgreen}{RGB}{175, 193, 36}
\usetikzlibrary{arrows,shapes,shadows,mindmap,fit,calc,shapes.misc,positioning,decorations.pathreplacing}

\tikzstyle{mybox} = [draw=blue!5, fill=blue!5, very thick,
    rectangle, rounded corners, inner sep=10pt, inner ysep=10pt]
\tikzstyle{fancytitle} =[fill=red, text=white]

\tikzset{%
  highlight/.style={rectangle,rounded corners,fill=green!15,draw,fill opacity=0.2,thick,inner sep=0pt}
}

\newcommand{\graphitemize}[2]{%
\begin{tikzpicture}[every node/.style={align=center}]  
  \node[minimum size=3.2cm,circle,fill=gray!20,font=\Large,outer sep=1cm,inner sep=.5cm](ce){#1};  
\foreach \gritem [count=\xi] in {#2}
{\global\let\maxgritem\xi}  
\foreach \gritem [count=\xi] in {#2}
{%
\pgfmathtruncatemacro{\angle}{360/\maxgritem*\xi + 20}
\edef\col{blue!20}
\node[circle,
     ultra thick,
     draw=white,
     fill opacity=.8,
     fill=\col,        
     minimum size=2.5cm] at (ce.\angle) {\gritem };}%
\end{tikzpicture}  
}%
\tikzstyle{level 1}=[level distance=3.5cm, sibling distance=3.5cm]
\tikzstyle{level 2}=[level distance=3.5cm, sibling distance=2cm]

\tikzstyle{bag} = [text width=4em, text centered]
\tikzstyle{end} = [circle, minimum width=3pt,fill, inner sep=0pt]

\newcommand{\errplot}{%
  \begin{tikzpicture}[trim axis left,trim axis right]
    \begin{axis}[y=-\baselineskip,
        scale only axis,
        width             = 6.5cm,
        enlarge y limits  = {abs=0.5},
        axis y line*      = middle,
        y axis line style = dashed,
        ytick             = \empty,
        axis x line*      = bottom
      ]
      \addplot+[only marks][error bars/.cd,x dir=both, x explicit]
        table [x=mean,y expr=\coordindex,x error=error]{\data};
    \end{axis}
  \end{tikzpicture}%
}


%% file: table1.tex
\pgfplotstableset{create on use/error/.style={
    create col/expr={\thisrow{uci}-\thisrow{mean}
    }
  }
}

\pgfplotstablegetrowsof{\data}
\let\numberofrows=\pgfplotsretval

\resizebox{1.0\linewidth}{!}{
\pgfplotstabletypeset[columns={name,error,mean,ci},
  every head row/.style={before row=\toprule,after row=\midrule},
  every last row/.style={after row=[3ex]\bottomrule},
  columns/name/.style={string type,column name=Name},
  columns/feature/.style={string type,column name=Parameter},
  columns/error/.style={
    column name={},
    assign cell content/.code={
    \ifnum\pgfplotstablerow=0
    \pgfkeyssetvalue{/pgfplots/table/@cell content}
    {\multirow{\numberofrows}{6.5cm}{\errplot}}%
    \else
    \pgfkeyssetvalue{/pgfplots/table/@cell content}{}%
    \fi
    }
  },
  columns/mean/.style={column name=Mean,fixed,fixed zerofill,dec sep align},
  columns/ci/.style={string type,column name=95\% CI},
  create on use/ci/.style={
    create col/assign/.code={\edef\value{(
    \noexpand\pgfmathprintnumber[showpos,fixed,fixed zerofill]{\thisrow{lci}} to \noexpand\pgfmathprintnumber[showpos,fixed,fixed zerofill]{\thisrow{uci}})}
      \pgfkeyslet{/pgfplots/table/create col/next content}\value
    }
  }
]{\data}
}